\documentclass[showpacs,twocolumn,amsmath,amssymb]{revtex4-1}
\usepackage{forloop,ifthen} 
\usepackage{multirow} 
\usepackage{amsmath} 
    \usepackage{graphicx}
    \usepackage{float}
    \usepackage{color}
    \usepackage{dcolumn}
    \usepackage{bm}
    \usepackage{xspace}
    \usepackage[colorlinks=true,     linkcolor={red!70!black},
    citecolor={green!50!black}, urlcolor={blue!50!black}, pdfborder={0 0 0}]{hyperref}
    \usepackage{calc}
    \newcommand{\suppress}[1]{}
    \newcommand{\CO}[1]{{#1}}
   
   \newcommand{\hrefBlank}[2]{}

   \usepackage[usenames,dvipsnames]{xcolor}
   \usepackage{ esint }
   \usepackage{ amssymb }
   \usepackage{graphicx}  
   \usepackage{fancybox}  
   
\begin{document}

\title{The Quantum Liouville Equation is non-Liouvillian}

\author{Dimitris Kakofengitis and Ole Steuernagel}

\affiliation{School of Physics, Astronomy and Mathematics, University of
Hertfordshire, Hatfield, AL10 9AB, UK}

\email{O.Steuernagel@herts.ac.uk}

\date{\today}

\begin{abstract}
  The Hamiltonian flow of a classical, time-independent, conservative system is
  incompressible, it is Liouvillian. The analog of Hamilton's equations of motion for a
  quantum-mechanical system is the quantum-Liouville equation. It is shown that its
  associated quantum flow in phase space, Wigner flow, is not incompressible. It gives
  rise to a quantum analog of classical Hamiltonian vector fields: the Wigner phase space
  velocity field~$\bm w$, the divergence of which can be unbounded. The loci of such
  unbounded divergence form lines in phase space which coincide with the lines of zero of
  the Wigner function. Along these lines exist characteristic pinch points which coincide
  with stagnation points of the Wigner flow.
\end{abstract}


\maketitle

\section{Introduction}

In classical phase space the coordinates $\bm r = \binom{\bm q}{\bm p}$ are position~${\bm
  q}$ and momentum~${\bm p}$ with the associated dynamics described by the Hamiltonian
velocity field $\bm v = \binom{\bm {\dot q}}{\bm {\dot p}}$ giving rise to a continuity
equation~$\frac{\partial}{\partial t} \rho(\bm r;t) + \bm \nabla \bullet \bm j(\bm r;t) =
\sigma(\bm r;t)$ for the movement of the classical probability density~$\rho$ and its
flow~$\bm j$. Because probability is locally conserved the source term $\sigma(\bm r;t)
=0$.

Famously, classical Hamiltonian vector fields for time-independent conservative systems are
\noindent\begin{eqnarray}
\text{divergence-free } \quad &\bm \nabla \bullet \bm v \; & =  0 ,
\label{eq:_Div_v_zero_classical}
\\
\text{or, the flow is incompressible } \quad &\frac{ D \rho }{Dt} & =  0 .
\label{eq:_Dtotal_zero_classical}
\end{eqnarray}
Eq.~(\ref{eq:_Dtotal_zero_classical}) follows from~(\ref{eq:_Div_v_zero_classical}) 
if the density is made up of ``carriers'', particles or charges (and their respective
probability distributions), such that the flow $\bm j$ can be decomposed into the product
\begin{equation}
 \bm j = \rho \bm v .
\label{eq:_j_rho_times_velocity}
\end{equation}
Then, the \emph{total derivative}~\cite{ComovingDerivativeNames} of~$\rho$ is $ \frac{ D
  \rho}{Dt} = - \rho \bm \nabla \bullet \bm v$.

In quantum mechanics, Wigner's quantum phase space-based distribution
function~$W$~\cite{Wigner_32,Hillery_PR84} obeys~\cite{Wigner_32} the, so-called, quantum
Liouville equation~\cite{Schleich_01}
\begin{equation} \frac{\partial W(\bm r;t) }{\partial t} + \bm \nabla \bullet \bm J(\bm
  r;t) = 0 \; ,
\label{eq:W_Continuity}
\end{equation}  
where~$\bm J$ is the Wigner flow~\cite{Ole_PRL13} of the system.

Here we establish that the quantum Liouville equation is typically non-Liouvillian, that
Wigner's phase space velocity~$\bm w$, the quantum analog of~$\bm v$, can have unbounded
divergence, and that the structure of the divergence of~$\bm w$ can help us to investigate
the phase space structure of a quantum system's dynamics.

We first review features of Wigner flow and introduce the concept of the Wigner phase
space velocity~$\bm w$~(\ref{eq:w}), in section~\ref{sec:WignerFlow}. We then consider
quantum systems for which the flow of~$\bm w$ is always incompressible (harmonic
oscillator), in section~\ref{sec:HarmWignerFlow}, incompressible for energy eigenstates
only (`squared' harmonic oscillator), in section~\ref{sec:KerrWignerFlow}, and generically
non-Liouvillian (anharmonic oscillator), in section~\ref{sec:AnHarmWignerFlow}, before we
conclude in section~\ref{sec:Conclusion}.

\section{Wigner Flow\label{sec:WignerFlow}}

From now on, we will only consider motion in one spatial dimension~$x$. In this case~$W$ is
a one-dimensional Fourier transform
\begin{equation}\label{eq:W}
  W_\varrho(x,p;t) \equiv \frac{1}{\pi \hbar} \int_{-\infty}^{\infty} dy \, 
  \varrho(x+y,x-y,t) \cdot e^{\frac{2i}{\hbar} p y} \; ,
\end{equation}
of the off-diagonal coherences~$\varrho(x+y,x-y,t)$ of the quantum mechanical density
matrix~ $\varrho$ which has the form~$\varrho = \Psi^*(x+y,t)\Psi(x-y,t)$ if the system is
in a pure state~$\Psi$ (star `*' denotes complex conjugation); $\hbar=h/(2\pi)$ is
Planck's constant rescaled.

$W$ is real valued but can be negative~\cite{Wigner_32} and therefore is a
quantum-mechanical `quasi-probability' function~\cite{Hillery_PR84,Schleich_01}.

For time-independent conservative systems such as a point mass~$M$ moving under the
influence of a potential~$U$, described by the Hamiltonian
\begin{equation}
  H(x,p) = \frac{p^2}{2M} + U(x) ,
\label{eq:_classical_Hamiltonian}
\end{equation}
where the potential~$U(x)$ can be Taylor-expanded (giving rise to finite forces only),
$\bm J$ of~(\ref{eq:W_Continuity}) has the explicit form~\cite{Wigner_32}
\begin{eqnarray}
{\bm J} =  \binom{ J_x }{ J_p } = 
 \begin{pmatrix} \frac{p}{M} W 
    \\ -\sum\limits_{l=0}^{\infty}{\frac{(i\hbar/2)^{2l}}{(2l+1)!}
\partial_p^{2l}W
\partial_x^{2l+1}U
} 
 \end{pmatrix}
\; .
\label{eq:FlowComponents}
\end{eqnarray}
Here, the notation~$\partial_x^l=\frac{\partial^l}{\partial x^l}$, etc., is used for
conciseness. Explicit reference to dependence on $\bm r$ and $t$ is now dropped.

Wigner flow's complicated form makes it non-trivial to work out its overall structure.

To characterize Wigner flow it is useful to determine its orientation winding
number~\cite{Ole_PRL13} (or Poincar\'e index)
\begin{eqnarray}
  \label{eq:WindingNumber}
  \omega({\cal L},t) =\frac{1}{2\pi} \varointctrclockwise_{\cal L} d
  \varphi \; . 
\end{eqnarray}
The Poincar\'e index $\omega$~tracks the orientation angle~$\varphi$ of the flow
vectors~$\bm J$ along continuous, closed, self-avoiding loops~$\cal L$ in phase
space. Because the components of the flow are continuous functions, $\omega$ is zero
except for the case when the loop contains stagnation points. In such a case a non-zero
value of $\omega$ can occur and this value is conserved unless the system's dynamics
transports a stagnation point across~$\cal L$~\cite{Ole_PRL13}.

When comparing Wigner flow with classical Hamiltonian flow, it is not unreasonable to
argue that the first order terms of Wigner flow~(\ref{eq:FlowComponents}) have classical
form
\begin{equation}
\binom{ J_x }{ J_p }  = \binom{ v_x W}{ - W \partial_x V } + 
\binom{ 0 }{ {\cal O}(\hbar^2) } \ ;
\label{eq:WFlow_mot_Classical}
\end{equation}
and therefore, whenever higher order quantum terms~${\cal O}(\hbar^2)$ are present, Wigner
flow cannot be Liouvillian~\cite{Zachos_book_05}. It turns out that for eigenstates of
Kerr oscillators (section~\ref{sec:KerrWignerFlow}, below) this is not correct though.

We note that, firstly, Wigner's function typically has areas of negative value which is
why classical probability arguments have to be used cautiously. Secondly, a clear
identification of the terms responsible for deviation from the classical case might be of
interest in its own right. And, thirdly, we have, so far, little intuition regarding the
behaviour of Wigner flow, and we show here that the divergence of its flow can be tied to
other physical phenomena, such as the formation of stagnation points of~$\bm J$ in phase
space.

To establish that in general quantum phase space flow is non-Liouvillian, let us cast it
into a form analogous to Eq.~(\ref{eq:_j_rho_times_velocity}), namely~$ {\bm J} = W {\bm
  w}$, and investigate the divergence of the \emph{Wigner phase space velocity}
\begin{equation}\label{eq:w}
\bm w = \frac{\bm J}{W} .
\end{equation}

According to Eq.~(\ref{eq:_Div_v_zero_classical}), to establish when Wigner flow is
Liouvillian, we determine when
\begin{equation}  \boldsymbol{\nabla} \bullet \bm w = 0 \; . 
\label{eq:Wigner_velocity_Liouvillian}
\end{equation} 

With $\boldsymbol{\nabla} \bullet \bm J = W \boldsymbol{\nabla} \bullet \bm w + \bm w
\bullet \boldsymbol{\nabla} W = -\partial_t W$ we have
\begin{equation}
 \boldsymbol{\nabla} \bullet \bm w = - \frac{ { \bm J}
\bullet \boldsymbol{\nabla} W  + W \partial_t W}{W^2} .
\label{eq:Wigner_velocity_Divergence}
\end{equation}

\section{Wigner Flow of Harmonic Oscillators \label{sec:HarmWignerFlow}}

\begin{figure}[t]
\centering
  \includegraphics[width=0.41\textwidth]{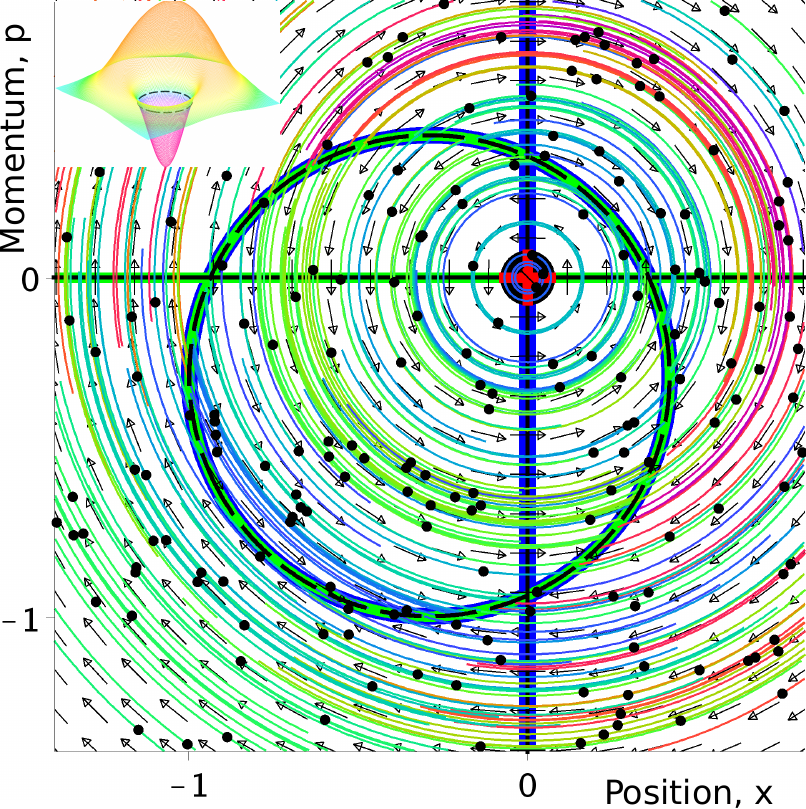}
  \caption{\CO{(Color online)} Normalized Wigner flow~$\bm J / |\bm J|$ with its
    streamlines for superposition state $\Psi=\cos(\frac{\pi}{3}) |0\rangle +
    \sin(\frac{\pi}{3}) e^{-i \frac{7}{4} \pi} |1\rangle $ of a harmonic
    oscillator; with~$M=1$,~$k=1$ and $\hbar=1$. The streamlines, which were randomly
    picked and coloured, are~\emph{circular}.  Despite quantum mechanical flow
    inversion~\cite{Ole_PRL13} for~$W<0$ (see inset), the harmonic oscillator's Wigner
    flow is always Liouvillian.  The black dashed line on top of the green circle depicts
    the line of zero of the Wigner function (see inset). The red cross at the origin marks
    the position of the flow's stagnation point, around the potential minimum, with
    Poincar\'e index~$\omega=1$.
    \label{fig:flow_harm}}
\end{figure}

For a harmonic potential $ U^\odot(x) = \frac{k}{2}x^2 $ with spring
constant~$k$, Wigner flow~(\ref{eq:FlowComponents}) has the `classical' form
\begin{equation} {\bm J}^\odot = \binom{J^\odot_{ x}}{J^\odot_{ p}} = W^\odot(x,p,t) \cdot \binom{
    \frac{p}{M} }{ - k x } \, .
  \label{eq:WF_harm}
\end{equation}
Inserting~(\ref{eq:WF_harm}) into~(\ref{eq:Wigner_velocity_Divergence}) yields $ \bm
\nabla \bullet \bm w^\odot = 0 $, always. A harmonic oscillator's quantum phase space flow
is always Liouvillian, see Fig.~\ref{fig:flow_harm}.

\section{Wigner Flow of the Kerr Oscillator\label{sec:KerrWignerFlow}}

An example of a system for which its energy eigenstates yield Liouvillian Wigner flow, but
its superposition states do not, is the `squared' harmonic oscillator, described by the Kerr
Hamiltonian
\begin{equation}
  {\cal \hat H_K} = \left( \frac{\hat p^2}{2M} + \frac{k}{2} \hat x^2 \right) + \Lambda^2 \left( \frac{\hat p^2}{2M} + \frac{k}{2} \hat x^2 \right)^2 \, .
  \label{eq_Kerr_Hamiltonian}
\end{equation}
The parameter $\Lambda$ parameterizes the system's (quantum-optical) Kerr--non-linearity,
$\Lambda \propto
\sqrt{\chi^{(3)}}$~\cite{Walls_Milburn_QuopBook,Osborn_JPA09,Manko_PSc10,Kirchmair_NAT13},
i.e. in field operator language~$ {\cal \hat H_K} = \left( a^\dagger a + \frac{1}{2}
\right) + \tilde \chi^{(3)} \left( a^\dagger a + \frac{1}{2} \right)^2$. The wavefunctions
of the harmonic oscillator are solutions to the Kerr Hamiltonian rendering the entire
system analytically solvable.

Note that ${\cal \hat H_K}$ contains products in $\hat x$ and $\hat p$, this implies that
the terms for the Wigner flow are not of the form~(\ref{eq:FlowComponents}). Instead, the
Wigner flow components can be determined using Moyal brackets~\cite{Zachos_book_05} and
are found to be of the form~\cite{Kerr_WignerFlow}

\begin{figure*}[t]
  \centering
   \includegraphics[width=0.8\textwidth,angle=0]{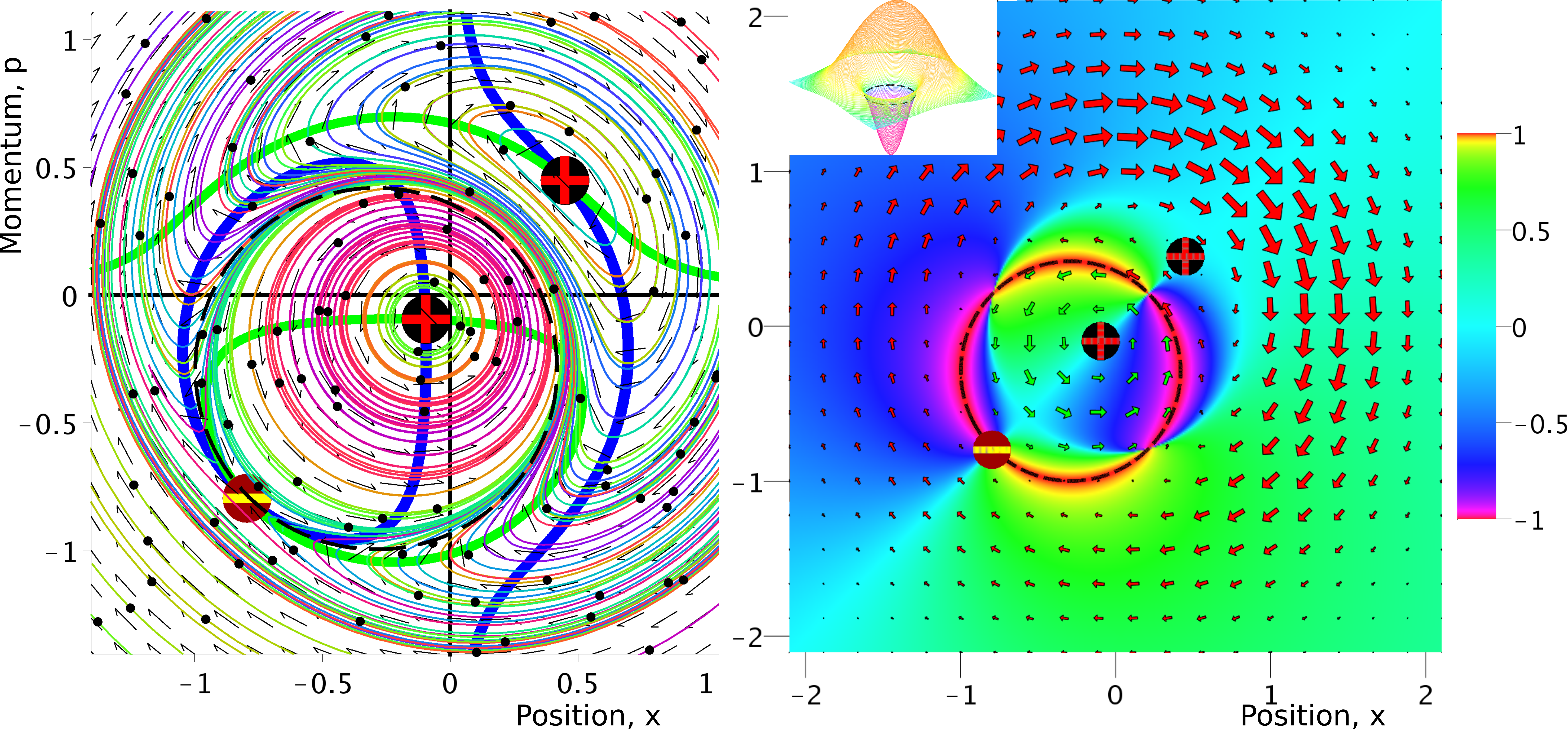}
   \caption{\CO{(Color online)} Behaviour of superposition state~$\Psi$ for the Kerr
     oscillator~(\ref{eq_Kerr_Hamiltonian}) with~$\Lambda = 2$, all other parameters as in
     Fig.~\ref{fig:flow_harm}.  Left: Streamlines of Wigner flow~$\bm J$ and its
     stagnation points.  Yellow minus signs mark stagnation points with Poincar\'e
     index~$\omega = - 1$. Right: $\frac{2}{\pi} \arctan (\bm \nabla \bullet \bm w)$ with
     Wigner flow~$\bm J$ superimposed. $\bm \nabla \bullet \bm w \neq 0$ almost
     everywhere: for superposition states quantum phase space flow of the Kerr system is
     non-Liouvillian. The black dashed line marks the zero of the Wigner function (see
     inset), at its location, according to Eq.~(\ref{eq:Wigner_velocity_Divergence}), the
     divergence of~$\bm w$ becomes unbounded (see right panel).
     \label{fig:flow_Kerr}}
\end{figure*}
\begin{widetext}
\begin{eqnarray} 
& J_x \; & = \left[ \Lambda^2 \left( -{ \frac {\hbar^2 p}{4{M}^{2}}} \partial_{x}^{2}
    + \left\{ { \frac {{p}^{3}}{{M}^{2}}}+ {\frac {k{x}^{2}p}{M}}
    \right\} -  {\frac {\hbar^2 k}{4 M}} \partial_{p}^ {2} p \right)+\left\{ \frac
    {p }{M} \right\} \right] W(x,p,t) \label{eq:J_x}
\\
\mbox{ and } 
\qquad & J_p\; & = \left[ \Lambda^2 \left( \frac{\hbar^2 k^2 x}{4} \partial_{p}^2 - \left\{
       {k}^{2}{x}^{3} + {\frac {kx{p }^{2}}{M}} \right\} +  \frac {\hbar^2 k}{4 M} \partial_{x}^2 x \right) - \left\{ k x \right\} \right] W(x,p,t) \; ,
\label{eq:J_p}
\end{eqnarray} 
\end{widetext}
where the curly brackets surround the classical terms. 
 All other terms (of order~${\cal O}(\hbar^2)$) are of quantum origin. 

 For symmetry reasons the quantum terms cancel for eigenstates but not otherwise and are
 responsible for the non-Liouvillian nature of quantum phase space flow of superposition
 states of the Kerr oscillator. For energy eigenstates,
 Eq.~(\ref{eq:Wigner_velocity_Divergence}) reads
\begin{equation}
 \boldsymbol{\nabla} \bullet \bm w = - \frac{ { \bm J}
\bullet \boldsymbol{\nabla} W }{W^2} .
\label{eq:Wigner_velocity_Divergence_Eigenstates}
\end{equation}
For eigenstates of the Kerr system, $\bm J$ is always perpendicular to $\bm \nabla W$ and
therefore quantum phase space flow is Liouvillian for its eigenstates.

For a superposition state (depicted in Fig.~\ref{fig:flow_Kerr}) Wigner flow is
non-Liouvillian and forms isolated flow stagnation points at the intersections of the
lines of vanishing $J_x$-component of the flow (thick green lines in
Fig.~\ref{fig:flow_Kerr}) with lines of vanishing $J_p$-component of the flow (thick blue
lines in Fig.~\ref{fig:flow_Kerr}). The flow's corresponding stagnation points are
depicted by red plus-signs, if their Poincar\'e index~$\omega=1$, and by yellow minus
signs, if their Poincar\'e index~$\omega=-1$.

\section{Wigner Flow of  Anharmonic Oscillators\label{sec:AnHarmWignerFlow}}

For Hamiltonians of the form~(\ref{eq:_classical_Hamiltonian}) with an anharmonic
potential~$U$ it is no longer true that the divergence of~$\bm w$ for eigenstates is
zero. In this case Wigner flow typically expands or compresses, i.e., is non-Liouvillian
always, almost everywhere in phase space. This can be understood from the previous
discussion of Eq.~(\ref{eq:WFlow_mot_Classical}). The quantum terms in Wigner flow yield
terms that break the incompressibility of classical phase space flow~\cite{Zachos_book_05}
and there are no symmetries, such as those for the eigenstates of the Kerr system, to
offset their influence.

\begin{figure*}[t]
  \centering
  \includegraphics[width=0.8\textwidth,angle=0]{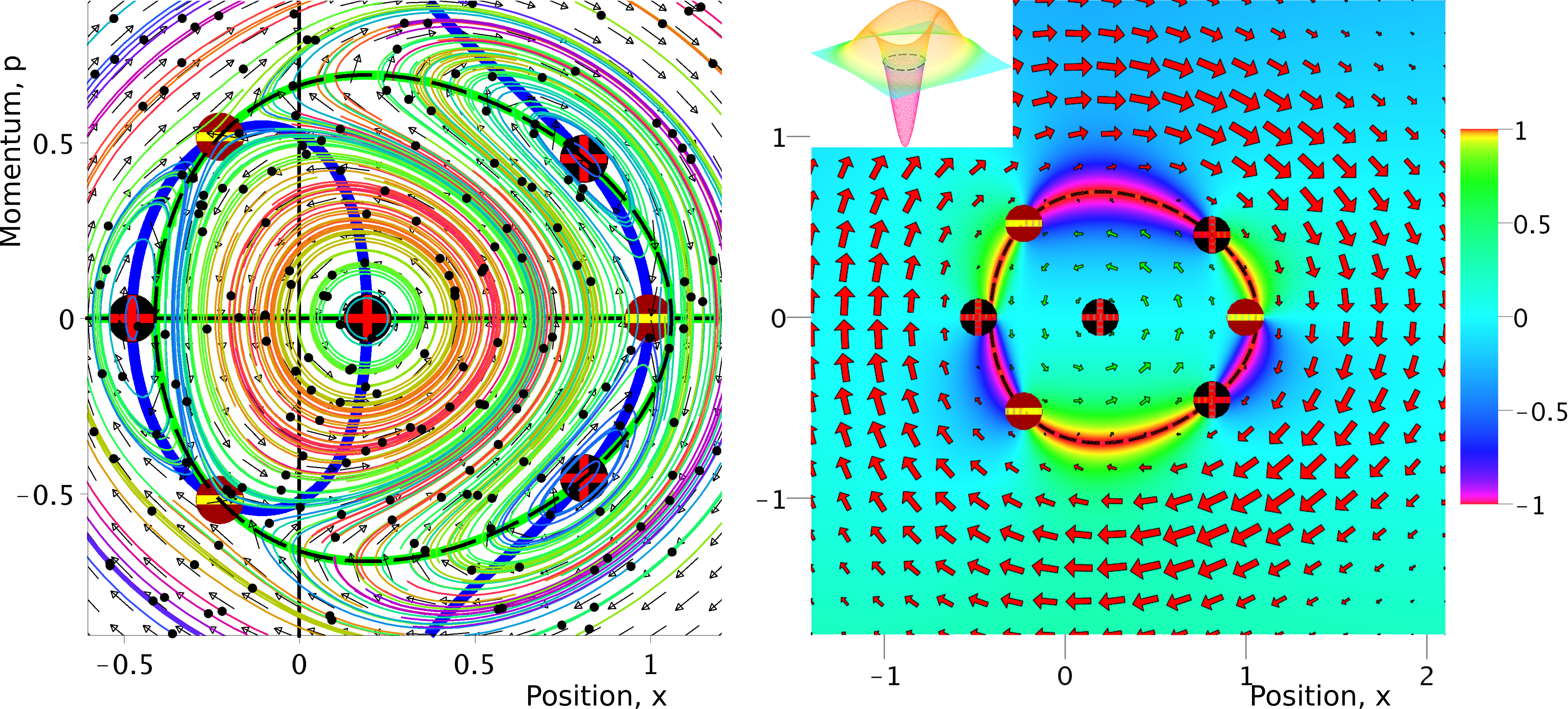}
  \caption{\CO{(Color online)} First excited energy eigenstate of the Morse oscillator
    with potential~$U(x)=8(1-\exp[-\frac{x}{4}])^2$ ($M=1$ and $\hbar=1$). Left:
    Streamlines of Wigner flow~$\bm J$ and its stagnation points (as in
    Figs.~\ref{fig:flow_harm} and~\ref{fig:flow_Kerr}). Right: $\frac{2}{\pi} \arctan (\bm
    \nabla \bullet \bm w)$ with $\bm J$ superimposed (Wigner function shown in inset;
    $J$-vectors are colored red for $W>0$ and green if the flow is inverted because
    $W<0$). For an anharmonic oscillator~(\ref{eq:_classical_Hamiltonian}) Wigner flow is
    non-Liouvillian ($\bm \nabla \bullet \bm w \neq 0$) always, almost everywhere.
    \label{fig:flow_Anharmonic}}
\end{figure*}

For illustration we show the Wigner flow portrait and the associated divergence map for
the first excited bound state of a Morse oscillator in Fig.~\ref{fig:flow_Anharmonic}.

In the case of mechanical quantum systems, described by Hamiltonians of the
form~(\ref{eq:_classical_Hamiltonian}), the (black dashed) line of zero of the Wigner
function, according to Eq.~(\ref{eq:WFlow_mot_Classical}), coincides with (thick green)
lines of zero of the $J_x$-component.  According to
Eq.~(\ref{eq:Wigner_velocity_Divergence}) this is the location where the divergence of the
Wigner phase space velocity~$\bm w$ becomes unbounded.  The (thick blue) lines of
vanishing $J_p$-component of the flow do typically not coincide with $J_x$-zero lines;
this leads to the formation of isolated stagnation points of the
flow~\cite{Ole_PRL13,Kakofengitis14} wherever (off the $x$-axis) blue and green lines
cross each other, see Fig.~\ref{fig:flow_Anharmonic}. In other words, when we follow the
line of unbounded divergence of~$\bm w$ we trace out the line where $J_x=0$. If such a
line crosses (off the $x$-axis) with a line where $J_p=0$, $\bm \nabla \bullet \bm w$
changes sign, this leads to the formation of the pinch-points of~$\bm \nabla \bullet \bm
w$ evident in the right panel of Fig.~\ref{fig:flow_Anharmonic}. Off the $x$-axis, these
pinch-points thus coincide with flow stagnation points.

\section{Conclusion\label{sec:Conclusion}}

We introduce the concept of the Wigner phase space velocity~$\bm w$.  We show that the
quantum-Liouville equation~(\ref{eq:W_Continuity}) is generically non-Liouvillian and
would better be called quantum-continuity equation. Only in the case of the harmonic
oscillator is the flow of the Wigner phase velocity divergence-free. Generically, for any
anharmonic quantum-mechanical oscillator, Wigner flow is non-Liouvillian and features
unbounded divergence. Field-oscillators of the Kerr type show intermediate behaviour in
that their eigenstates feature Liouvillian flow, but their coherent superpositions do not.
In anharmonic quantum-\emph{mechanical} systems~(\ref{eq:_classical_Hamiltonian}) the
(off-axis) pinch-points of unbounded divergence of Wigner's phase space velocity~$\bm w$
coincide with the stagnation points of Wigner flow~$\bm J$.

\bibliography{NonLiouvillian_WignerFlow.bbl}

\end{document}